# Momentum in the dynamics of variable-mass systems: classical and relativistic case

Janusz Wolny* and Radosław Strzałka

*Faculty of Physics and Applied Computer Science, AGH University of Science and Technology, Krakow, Poland*

Abstract

We discuss a role of a momentum vector in the description of dynamics of systems with variable mass, and show some ambiguity in expressing the 2nd Newton's law of dynamics in terms of momentum change in time for variable-mass systems. A simple expression that the time-derivative of the momentum of the body with variable mass is equal to the net external force is not always true (only if a special frame of reference is assumed). In basic textbooks and multiple lecture notes the correct equation of motion for a variable-mass system (including relative velocities of the masses entering or leaving the body) is not sufficiently well discussed, leading to some problems with understanding the dynamics of these systems among students. We also show how the equation of motion in classical case (in translational motion) can be easily expanded to the relativistic case and discuss a motion of a relativistic rocket. For the non-relativistic case also a rotational motion is discussed. It is of course true, that most of the good literature treats the problem accurately, but some of the commonly used textbooks do not. The purpose of our letter is to pay attention to the problem of dynamics of variable-mass systems and show yet another perspective of the subject.

I. INTRODUCTION

Momentum is a quantity commonly used to describe the state of the system's motion. Frequently, the momentum (**p**) is used to present dynamical equations in the form that the momentum's first derivative equals the net force (**F**) acting on the body, i.e. $\frac{d\mathbf{p}}{dt} = \mathbf{F}$. This elegant way of writing the second Newton's law of dynamics is fully equivalent to the classic expression $m\mathbf{a} = \mathbf{F}$ ($m$-mass, **a**-acceleration), but only for systems with a constant mass. The mindless use of the momentum to write equations of motion for a variable-mass system is a common behavior and can lead to serious mistakes reproduced in many lectures and in many textbooks in physics.[1] Discussion on this subject is often explained unclear. Our publication will derive the equation of motion of a variable-mass body based on the classical principles of dynamics. In a very simple way a classical equation will be generalized to the relativistic case. The conditions will be strictly defined when it can be reduced to the simple formula using a momentum, independently on the inertial frame of reference. The combination of a generalized equations of motion for systems with variable mass and the relativistic principle of mass and energy equivalence ($E = mc^2$) leads to equations of dynamics in special relativity.

The motivation for writing this letter is the universal manner of writing the second law of dynamics in the form of a time-derivative of momentum, i.e. $\frac{d\mathbf{p}}{dt} = \mathbf{F}$. This record is valid only for systems with a constant mass, when it is equivalent to the second principle of Newton's dynamics, i.e.: $m\mathbf{a} = m\frac{d\mathbf{v}}{dt} = \mathbf{F}$.[4] For systems with variable mass, the two above formulae are contradictory. The latter equation, $m\mathbf{a} = \mathbf{F}$, is valid, and it remains identical in all inertial reference systems. The expression with momentum can be valid only for strictly defined systems, if the conditions of the task allow such a choice.

II. CLASSICAL MECHANICAL SYSTEM WITH VARIABLE MASS

The simplest case of a variable-mass system is a rocket engine. This is an example of the interaction of only two bodies: a rocket and fuel ejected from it. Let us denote by $\mu \equiv \frac{dm}{dt}$ a mass change of a rocket (in kilograms per second, kg/s; note, that $\mu < 0$, since the mass, $m$, of a rocket decreases; a fuel combustion rate is then $-\frac{dm}{dt}$), by $\mathbf{v}$ the velocity vector of the rocket, and by $\mathbf{u}$ the velocity vector of ejected fuel in its own frame of reference.

We will now use the third law of dynamics, i.e. the equality of forces of mutual interaction of bodies, to describe dynamics of the rocket. During a short time $\tau$ the portion of fuel mass $\Delta m = -\mu\tau$ gets velocity **u**, so it must move in accelerated motion with acceleration vector $\mathbf{a}_{gas} = \frac{\mathbf{u}}{\tau}$, under force $\mathbf{F}_{gas}$, so that $\mathbf{F}_{gas} = \Delta m \mathbf{a}_{gas} = -\mu\tau\frac{\mathbf{u}}{\tau} = -\mu\mathbf{u}$. According to the third law of dynamics in the two-body interaction, which is the system rocket–exhaust, the recoil force is opposite to the net force of gases, and it causes the thrust of the rocket, i.e. $\mathbf{F}_{thrust} = -\mathbf{F}_{gas} = \mu\mathbf{u}$. This equation is correct if the whole interaction takes place in a two-body system (rocket–exhaust gases). The origin of this interaction is the pressure of the hot gas. In the case of partial interaction (e.g. part of the gases escapes through engine leaks) corrections expressed by appropriate coefficients of less than one should be introduced. Finally, for an ideal rocket engine, the simple use of dynamics laws leads to the equation:

$$m\frac{d\mathbf{v}}{dt} = \mathbf{F}_{ext} + \mu\mathbf{u}, \qquad (1)$$

known as the Meshchersky equation.[5,6] $\mathbf{F}_{ext}$ is the external force (e.g. gravity or air resistance). The solution of the Meshchersky equation, assuming a constant $\mu$ and **u**, and no external forces, is the Tsiolkovsky formula for the final speed of the rocket (and assuming the initial speed of a rocket equal to 0):

$$v = u \ln\frac{m_0}{m}, \qquad (2)$$

where $m_0$ is the starting mass of the rocket, $m$ is the instantaneous mass, $m = m(t) = m_0 + \mu t, (\mu < 0)$.

The formula in Eq. (1) can be extended to the case, where several different sources of thrust are present, including the mass of air taken from outside, used for fuel combustion:

$$m\frac{d\mathbf{v}}{dt} = \mathbf{F}_{ext} + \sum_{i=1}^{n}\mu_i\mathbf{u}_i. \qquad (3)$$

An example of the application of the Eq. (3) may be a jet plane flying at velocity **v** and sucking in air from the outside in the amount of $\mu_1 > 0$ kg/s with the relative velocity $\mathbf{u}_1 = -\mathbf{v}$. The air is mixed with the aviation fuel, which is burned at the rate $\mu_{fuel} > 0$ kg/s. After combustion of the mixture of fuel with air in the engine's chamber it is ejected through the jet engine nozzles in the form of exhaust gases in the amount $\mu_2 = (\mu_1 + \mu_{\text{fuel}})$ kg/s at the relative velocity $\mathbf{u}_2$ with the opposite direction to $\mathbf{v}$. The equation of motion of such an aircraft with an ideal jet engine is then following:

$$m\frac{d\mathbf{v}}{dt} = \mathbf{F}_{\text{drag}} + \mu_1 \mathbf{u}_1 + \mu_2 \mathbf{u}_2 = \mathbf{F}_{\text{drag}} - \mu_1 \mathbf{v} + (\mu_1 + \mu_{\text{fuel}})\mathbf{u}_2, \qquad (4)$$

where $m$ is the current mass of the aircraft with fuel, $m = m_0 - \mu_2 t, (\mu_2 > 0)$, and $\mathbf{F}_{\text{drag}}$ is the speed dependent air resistance acting on the aircraft (playing a role of the external force). This approach to the problem is logical and very simple, and can be understood even for average student. In many textbooks, tasks of this type are very often inaccurately explained, which may lead to random, often erroneous final results.

```
EXERCISE 1. Thrust of the jet aircraft.

A jet aircraft moves at constant speed 250 m/s, which is also a speed of sucked air
into the engine. In each second the mixture of 75 kg of air and 3 kg of aviation
fuel is combusted in the engine, and the exhaust gases are ejected with the speed
500 m/s. What is the total thrust of the jet?

1. Example of wrong solution (frequently observed in textbooks).
```
An incorrect assumption is that the total mass $m_{\text{tot}} = 75 + 3 = 78$ kg of the gases is ejected with a relative speed $v_{\text{rel}} = 500 - 250 = 250$ m/s. The thrust is then wrongly assumed to be: $F_{\text{thrust}} = m_{\text{tot}} v_{\text{rel}} = 78 \cdot 250 = 19{,}500$ N.

```
2. Correct solution.
```
The correct solution is obtained, if the Eq. (4) is used for a thrust force, i.e. $F_{\text{thrust}} = -\mu_1 v + (\mu_1 + \mu_{\text{fuel}}) u_2$. We have: $\mu_1 = 75$ kg/s, $\mu_{\text{fuel}} = 3$ kg/s, $\mu_1 + \mu_{\text{fuel}} = 78$ kg/s, $v = 250$ m/s, and $u_2 = 500$ m/s. Thus, the total thrust is: $F_{\text{thrust}} = -75 \cdot 250 + 78 \cdot 500 = 20{,}250$ N.

By introducing the velocity of exhaust gases with respect to the laboratory (LAB) frame of reference (resting reference frame), $\mathbf{v}_1$, so that $\mathbf{u} = \mathbf{v}_1 - \mathbf{v}$, we get after the transformations of the Eq. (1):

$$m\frac{d\mathbf{v}}{dt} + \mathbf{v}\frac{dm}{dt} = \mathbf{F}_{\text{ext}} + \mu \mathbf{v}_1. \qquad (5)$$

After entering the momentum, $\mathbf{p} = m\mathbf{v}$, into the Eq. (5), we get:

$$\frac{d\mathbf{p}}{dt} = \mathbf{F}_{ext} + \mu\mathbf{v}_1. \tag{6}$$

From the Eq. (6), we can go to the law of conservation of momentum. The momentum of the body (also of the variable mass) is conserved provided that the sum of external forces and recoil is equal to zero: $\mathbf{F}_{ext} + \mu\mathbf{v}_1 = 0$. The appearance of velocity $\mathbf{v}_1$ limits the applicability of the law to only selected systems. The law of conservation of momentum for the rocket with variable mass, thus, loses its universality and becomes a special case. Of course, you can take all the bodies interacting with each other (e.g. a rocket with exhaust gases is a system with a constant mass as a whole), then the forces of mutual interaction between all bodies are balanced. For such an overall system, the conservation principle of the momentum remains, as commonly known (e.g. for the center of mass of the system). However, the applicability of the law is still limited. The law of energy conservation must be included. For a two-body system, the solvability is only in two extreme cases, see elastic collisions (we assume full mechanical energy conservation) or completely inelastic collisions (we abandon the law of energy conservation, but we assume that all components will merge to one body). These are only hypothetical cases and often have little to do with the actual course of the phenomenon.

Equation (6) is a correct form of the variable-mass equation of motion with the use of momentum. Only in the reference frame of the exhaust gas, if it is also an inertial frame, the velocity $\mathbf{v}_1$ takes 0 and the Eq. (6) takes the form:

$$\frac{d\mathbf{p}}{dt} = \mathbf{F}_{ext}. \tag{7}$$

The Equation (7) is the simplest, but not always true, and also easily leads to the law of conservation of momentum. Using it requires the use of an exhaust gas's own reference frame, which is often a non-inertial frame (e.g. associated to a rocket). In addition, in the case of the generalized Eq. (3), the common own reference frame for bodies with several different velocities, may not exist.

A good example illustrating the above issue is the scheme: a barge flowing on the water and the sand falling on or from the barge. If the sand falls on the barge from the belt conveyor moving with a constant speed $\mathbf{v}_1$ against a LAB reference frame, the formula in Eq. (7) written in the sand's own reference frame can be used. However, if the sand spills out of the barge (e.g. it is thrown into the water by a conveyor through the stern), the Eq. (7) is not true, since the sand's own reference frame is also the non-inertial frame of the barge itself and the fictitious forces must be introduced.[9]

III. VARIABLE-MASS SYSTEM: RELATIVISTIC CASE

Another case is the use of the Eq. (7) in relativity theory. This is possible due to the zero initial speed of the growing part of the relativistic mass. It is, however, again not the most general formula. Let us now write the formula in Eq. (6) in the relativistic case. In the relativistic case we assume the equivalence of mass and energy written as $E = mc^2$, where $m$ is a speed-dependent relativistic mass, $m = m(\mathbf{v})$. Under the force $\mathbf{F}_{ext}$ the power transferred to the system is $\frac{dE}{dt} = \mathbf{F}_{ext} \cdot \mathbf{v}$, therefore the rate of change of the relativistic mass is given by the formula

$$\mu \equiv \frac{dm}{dt} = \frac{d}{dt}\left(\frac{E}{c^2}\right) = \frac{\mathbf{F}_{ext} \cdot \mathbf{v}}{c^2}. \tag{8}$$

By inserting the formula in Eq. (8) to Eq. (1), after taking $\mathbf{u} = -\mathbf{v}$ ($\mathbf{v}_1 = 0$, where $\mathbf{v}_1$ is the velocity vector of the relativistically growing mass in the LAB system, which is 0), we obtain a commonly known formula for the acceleration of the relativistic system written in the form:

$$m\frac{d\mathbf{v}}{dt} = \mathbf{F}_{ext} - \frac{(\mathbf{F}_{ext} \cdot \mathbf{v})}{c^2}\mathbf{v}. \tag{9}$$

Formula in Eq. (9) is a relativistic equivalent of the classical formula for a dynamics of the variable-mass system – Eq. (1). Writing Eq. (9) in the special case of an own frame of reference, i.e. when $\mathbf{v}_1 = 0$, we get

$$\frac{d\mathbf{p}}{dt} = \mathbf{F}_{ext}. \tag{10}$$

Equations (9) and (10) are equivalent to each other if the own frame of reference is assumed (compare with Ref. 10), and they are the result from the corresponding classic relations – Eqs. (1) and (6).

It must be remembered, however, that the Eq. (10) is valid only in the specific case described, when the total mass of a single object is considered. This case is equivalent to the above-discussed classical example of throwing sand from above on a moving barge, where the equation expressed by the momentum also applies. The equation expressed by the acceleration $\left(\mathbf{a} = \frac{d\mathbf{v}}{dt}\right)$, i.e. the Eq. (9), is completely equivalent to the latter. Many authors of textbooks draw the conclusion here about the superiority of the momentum- over acceleration-based dynamical equations.[11] This is an incorrect and erroneous conclusion, because it is only met for a specific situation.

We can additionally assume that the relativistic mass is expressed by the formula $m(\mathbf{v}) = m_0 \gamma$, where $m_0$ is the rest mass, and $\gamma = \gamma(\mathbf{v})$ is the relativistic factor, and calculate the expression for the change of relativistic mass $\mu = \frac{dm}{dt}$ (keeping $m_0 = const(t)$). After inserting it into Eq. (1), and still taking $\mathbf{u} = -\mathbf{v}$ ($\mathbf{v}_1 = 0$), and after simple calculations we get the relativistic equation of motion

$$m_0 \frac{dv}{dt} = \frac{F_{ext}^{\parallel}}{\gamma^3}, \tag{11}$$

or, after dividing by $m_0$, and denoting $a = \frac{dv}{dt}$ and $a_0 = \frac{F_{ext}^{\parallel}}{m_0}$,

$$a = \frac{a_0}{\gamma^3}, \tag{12}$$

where $F_{ext}^{\parallel}$ is the component of the force taken in the direction parallel to the velocity vector of the reference frame, i.e. $F_{ext}^{\parallel} = \left(\mathbf{F}_{ext} \cdot \frac{\mathbf{v}}{v}\right)$, vector $a$ is the acceleration in the LAB frame in the parallel direction, and vector $a_0$ – the acceleration in the own reference frame, also along vector $\mathbf{v}$. Equation (12) is usually derived directly from the Lorentz transformation, which additionally confirms the validity of the classical equation of motion for systems with variable mass.

```
EXERCISE 2. Derivation of the Eqs. (11) and (12)
```
Assume the relativistic mass $m(\mathbf{v}) = \gamma m_0$ with $\gamma(\mathbf{v}) \equiv \frac{1}{\left(1-\left(\frac{v}{c}\right)^2\right)^{1/2}}$, and $m_0 = const(t)$, and starting from a general formula in Eq. (1) derive Eq. (12) for the relativistic transformation of acceleration.

```
SOLUTION.
1) First we will derive the formula for the change of relativistic mass:
```

$$\mu = \frac{dm}{dt} = \frac{d}{dt}(\gamma m_0) = m_0 \frac{d\gamma}{dt} = m_0 \frac{1}{\left(1-\left(\frac{v}{c}\right)^2\right)^{3/2}} \frac{\mathbf{v}}{c^2} \cdot \frac{d\mathbf{v}}{dt} = m_0 \gamma^3 \frac{\mathbf{v}}{c^2} \cdot \frac{d\mathbf{v}}{dt}.$$

2) Now we can insert the above term to Eq. (1), taking $m = \gamma m_0$, and $\mathbf{u} = -\mathbf{v}$. We must assume a parallel component of the external force $F_{ext}^{\parallel}$, calculated along direction of the vector $\mathbf{v}$ (velocity of the reference frame), since we are interested in the relativistic transformation of the parallel components of the acceleration vector. We get:

$$\gamma m_0 \frac{dv}{dt} = F_{ext}^{\parallel} - \left(m_0 \gamma^3 \frac{v}{c^2}\frac{dv}{dt}\right)v \Leftrightarrow \gamma m_0 \frac{dv}{dt}\left(1+\gamma^2 \frac{v^2}{c^2}\right) = F_{ext}^{\parallel} \Leftrightarrow \gamma^3 m_0 \frac{dv}{dt} = F_{ext}^{\parallel} \Leftrightarrow \frac{dv}{dt} = \frac{1}{\gamma^3}\frac{F_{ext}^{\parallel}}{m_0}.$$

If we now define the parallel component of the acceleration vector in the own reference frame as $a_0 = \frac{F_{ext}^{\parallel}}{m_0}$ and the same vector in the LAB system as $a = \frac{dv}{dt}$, we finally get Eq. (12).

If we allow the change of the rest mass of the object (e.g. in the case of a relativistic rocket) then also Eq. (10) is not invariant and must be modified accordingly. Let us now assume $m_0 = m_0(t)$ and modify the result of calculation for the mass change in the Exercise 2, i.e. $\mu = \frac{dm}{dt} = \frac{d}{dt}[\gamma(\mathbf{v})m_0(t)]$. Vector $\mathbf{v}$ is the velocity of the rocket (it is also a vector of the moving reference system measured against LAB). The Equation (6) written in the relativistic case of the variable rest mass gets

$$\frac{d\mathbf{p}}{dt} = \mathbf{F}_{ext} + \gamma \mu_0 \mathbf{v}_2 + \gamma^3 m_0 \frac{\mathbf{v}}{c^2}\cdot \frac{d\mathbf{v}}{dt}\mathbf{v}_2, \qquad (13)$$

where $\mu_0 = \frac{dm_0}{dt}$, and $\mathbf{v}_2$ is the velocity vector of the ejected "relativistic" gases (in the LAB reference frame). After introducing a relativistic momentum, $\mathbf{p} = \gamma m_0 \mathbf{v}$, and assuming collinear vectors $\mathbf{v}_2$ and $\mathbf{v}$ (measured in the LAB system), we can use the relativistic velocity-addition formula, $u = \frac{v_2 - v}{1-\frac{v}{c^2}v_2}$, to write Eq. (5) for the relativistic case

$$m_0 \frac{dv}{dt} = \frac{F_{ext}^{\parallel}}{\gamma^3} + \frac{1}{\gamma^2}\mu_0 u. \qquad (14)$$

We see that taking $\mu_0 = 0$ (considering a constant rest mass) we come back to the Eq. (11).

Again, we see that the equation of motion written in the momentum domain is completely equivalent to the one written in the acceleration domain. Saying that Eq. (10) or Eq. (13) are more general than the Eqs. (9) or (14), in the relativistic case most particularly, is wrong. Because momentum is not a good dynamic variable, the law of conservation of momentum in the relativistic case is replaced by the law of four-momentum, resulting from the unique metric used in the Minkowski space.

IV. VARIABLE MASS IN ROTARY MOTION

Let us now focus on the case of rotary motion. For this type of motion, we define an angular momentum, $\mathbf{L} = \mathbf{r} \times \mathbf{p}$, to describe the dynamics of the system. The simple transformations lead to the equation of dynamics:

$$\frac{d\mathbf{L}}{dt} = \mathbf{r} \times (\mathbf{F}_{ext} + \mu \mathbf{v}_1). \tag{15}$$

The Equation (15) leads to the law of conservation of angular momentum under many assumptions similar to those discussed previously. You can also use the equation of motion for the mass point with variable mass $m$ in the form:

$$mr^2 \frac{d\boldsymbol{\omega}}{dt} = \mathbf{r} \times (\mathbf{F}_{ext} + \mu \boldsymbol{u}), \tag{16}$$

where $\boldsymbol{\omega}$ is the angular velocity vector.

V. CONCLUSIONS

The publication discusses the dynamics equations of variable-mass systems known from the literature. Large confusions are introduced by these equations, which on the one hand simplify the notation and calculations, but on the other hand, are only true after fulfilling a number of specific assumptions. Often the assumptions needed are ignored and the simplified equation of dynamics is used inappropriately. In particular, the momentum of the moving body is a good example of a physical quantity which may cause the trouble. Equations written using the momentum are not invariant after transformation to another frame of reference (also: the inertial frame). This trap often leads to incorrect use of dynamics equations for systems with variable mass. The safest is to use the Eq. (1) or the Eq. (3) in the generalized form. These equations are a simple result of a direct use of the Newton's laws of dynamics. It is safer to use interactions between bodies, and by this the 3$^{rd}$ law of dynamics, and to avoid the conservations laws (of momentum, angular momentum, or energy) that are only met in specific, hypothetical conditions. This approach allows the use of classic formulae to describe also the dynamics of relativistic systems. Because momentum is not a good dynamic quantity in the relativistic theory, the law of conservation of momentum is replaced by the law of conservation of the four-momentum, resulting from the unique metric used in the Minkowski space.

In conclusion, it should be noted that momentum is not a good dynamic variable and the elegant way of writing the equations of motion using momentum is limited to specific reference frames. This is particularly evident for systems with variable mass. In turn, taking into account the two-bodies interactions in classic Newton's equations of dynamics, we obtain completely correct and very clearly written equations of dynamics, which successfully also describe relativistic phenomena. Each of the arguments quoted here is widely known, but they are often incorrectly used or explained, especially in the education process of the new generation of physicists and engineers.

---

\* correspondence to: <wolny@fis.agh.edu.pl>

[1] In many textbooks commonly used in the didactics of physics the formulations or examples that could lead a reader to incorrect interpretations of the used formulas appear. This is even more true for lecture notes or materials unpublished. The problem is raised in literature (including scientific papers) by some authors (see for instance Refs. 2,3). However, we claim, that the topic still needs attention.

[2] Daniel Kleppner and Robert J. Kolenkow, *An Introduction to Mechanics* (McGraw-Hill, New York, 1973).

[3] Angel R. Plastino and Juan C. Muzzio, "On the use and abuse of Newton's second law for variable mass problems", Celestial Mechanics and Dynamical Astronomy **53**, 227–232 (1992).

[4] Wikipedia: https://en.wikipedia.org, entry: "Newton's laws of motion". In paragraph "Newton's second law" authors suggest (based on three references [17][18][19]) that formulas $\mathbf{F} = \frac{d\mathbf{p}}{dt}$ and $\mathbf{F} = m\mathbf{a}$ are exactly equivalent, `Since Newton's second law is valid only for constant-mass systems (…)'. This is, however, only true in classical mechanics. If relativistic case is considered, the momentum becomes not a suitable variable in dynamics, as we discuss in later parts of this work.

[5] Ivan Vsevolodovich Meshchersky (1859-1935) – Russian mathematician and physicist, known for his works on mechanics. In 1893 he described motion of a variable-mass point by the abovementioned equation, followed by a detailed description in his Master Thesis *The dynamics of a point of variable mass* published in 1897, and extension of the equations of motion in general case.

[6] The Equation (1) appears of course in many good handbooks and publications (mostly older ones, see for instance Refs. 7,4, and further refs. therein), however, many commonly used ones do not discuss the problem of variable-mass systems in a proper way, also at the very basic level, and do not even mention the Meshchersky's equation as a correct equation of motion for systems with variable mass (see for instance Ref. 8). This leaves the reader in the belief that the equation $\frac{d\mathbf{p}}{dt} = \mathbf{F}$ is more general than the equation $m\mathbf{a} = \mathbf{F}$.

[7] Arnold Sommerfeld, "Mechanics", in Lectures on Theoretical Physics, Vol. I (New York, 1952).

[8] David Halliday, Robert Resnick, and Jearl Walker, *Fundamentals of physics,* 9th edition (John Wiley & Sons, Inc., 2011) – or older/newer edition.

[9] Richard P. Feynman, Robert B. Leighton, and Matthew Sands, *The Feynman Lectures on Physics*, online access: http://www.feynmanlectures.caltech.edu/. Authors in the chapter 15-8 make a brilliant comment on the law of conservation of momentum in special relativity. Quote: `(…) <u>if action and reaction are still equal</u> (which they may not be in detail, but are in the long run), there will be

conservation of momentum in the same way as before (…)'. (Authors of course assume the relativistic definition of momentum.) In other words, only in special reference of frame, where the III Newton's law is met, we can use formula in Eq. (7) or equivalent Eq. (10) to show the conservation of momentum in the absence of external forces.

[10] John R. Taylor, *Classical Mechanics*, (University Science Books, 2005) – or older/newer edition. See chapter 15, Eq. (15.100) and Exercise 15.79, where two Eqs. (9) and (10) are (correctly) considered equivalent.

[11] For example: in chapter 37-11 of Ref. 8 authors suggest that by introducing a new definition of momentum in the relativistic case the law of conversation of momentum still holds in different inertial frames. We show that Eq. (10) with no external forces truly presents a law of conservation of momentum only if the special case of the own reference frame is considered. Moreover, if this condition is fulfilled, the two expressions for the second Newton's law (with acceleration and momentum) are equivalent.